\documentclass{aa}  

\usepackage{graphicx}
\usepackage{txfonts}
\usepackage{multirow}
\usepackage{natbib}
\newcommand{\masyr}{${\rm mas\, yr^{-1}}$}

\begin{document}

   \title{Astrometry with MCAO: HST-GeMS proper motions in the globular cluster NGC~6681}

   \subtitle{}

   \author{D. Massari\inst{1,2}
          \and
          G. Fiorentino\inst{1}
          \and
          A. McConnachie\inst{3}
          \and
          A. Bellini\inst{4}
          \and
          E. Tolstoy\inst{2}
          \and
          P. Turri\inst{5,3}
          \and
          D. Andersen\inst{3}
          \and
          G. Bono\inst{6,7}
          \and
          P.B. Stetson\inst{3}
          \and
          J.-P. Veran\inst{3}
          }

   \institute{INAF-Osservatorio Astronomico di Bologna, via Ranzani 1, 40127, Bologna, Italy\\
              \email{davide.massari@oabo.inaf.it}
         \and
            University of Groningen, Kapteyn Astronomical Institute, NL-9747 AD Groningen, Netherlands
         \and   
            Herzberg Astronomy and Astrophysics, National Research Council Canada, 5071 West Saanich Road, Victoria, BC V9E 2E7, Canada
         \and
            Space Telescope Science Institute, 3700 San Martin Drive, Baltimore, MD 21218, USA\\
          \and   
            Department of Physics and Astronomy, University of Victoria, 3800 Finnerty Road, Victoria, BC V8P 5C2, Canada
         \and   
            Dipartimento di Fisica, Universit\`{a} di Roma Tor Vergata, Via della Ricerca Scientifca 1, 00133 Roma, Italy
         \and
            INAF-Osservatorio Astronomico di Roma, Via Frascati 33, 00040 Monteporzio Catone, Italy
         \and
            INAF–Osservatorio Astronomico di Capodimonte, Via Moiariello 16, 80131 Napoli, Italy        
             }

   \date{Received July 18, 2016; accepted September 19, 2016}

 
  \abstract
   {}
   {For the first time the astrometric capabilities of the Multi-Conjugate Adaptive Optics (MCAO) facility GeMS with the GSAOI camera 
   on Gemini-South are tested to quantify the accuracy in determining stellar proper motions in the Galactic globular cluster NGC~6681. }
   {Proper motions from HST/ACS for a sample of its stars are already available, and this allows us to 
   construct a distortion-free reference at the epoch of GeMS observations that is used to measure and correct the temporally changing 
   distortions for each GeMS exposure. 
   In this way, we are able to compare the corrected GeMS images with a first-epoch of HST/ACS images
   to recover the relative proper motion of the Sagittarius dwarf spheroidal galaxy with respect to 
   NGC~6681.}
   { We find this to be $(\mu_{\alpha}\cos\delta, \mu_{\delta}) = (4.09,-3.41)$ \masyr, which matches previous
   HST/ACS measurements with a very good accuracy of 0.03 \masyr and with a comparable precision (r.m.s of 0.43 \masyr).}
   {This study successfully demonstrates that high-quality proper motions can be measured for quite large fields of view 
   ($85$\arcsec$\times85$\arcsec) with MCAO-assisted, ground-based cameras and provides a first, successful test of the performances of GeMS on 
   multi-epoch data. }


   \maketitle
%

\section{Introduction}

Proper motions (PMs) are an extremely powerful tool to investigate the kinematics and dynamics of stellar systems.
However, since their size depends on the distance of the objects under study and on the temporal baseline 
between the observations, PMs can be very small and difficult to measure. For example an object moving at 100 km/s 
at a distance of 100 kpc, has a PM of only $\sim0.2$ \masyr. 
This is why the internal kinematics of stellar systems have only been studied for the closest of them, i.e.
Galactic Globular Clusters (GCs).
The most reliable examples available in literature (see e.g. \citealt{mcl06, avdm10, mcnamara12, bellini14, watkins15}) exploit
the exceptional astrometric performance of the {\it Hubble Space Telescope} (HST), whose diffraction-limited PSF and geometric distortions
proved to be well determined and extremely stable over more than 20 years of operations (e.g. \citealt{anderson07, bellini11}).

Recently, diffraction limited observations have also been possible for ground-based telescopes 
and over quite large fields of view (FoV), thanks to the Multi-Conjugate Adaptive Optics (MCAO) technique (\citealt{ragazzoni00}).
This was first successfully tested on sky with the Multi-conjugate Adaptive optics Demonstrator (MAD, \citealt{marchetti08})
at the Very Large Telescope. 
Because of the size of the telescope, ground-based diffraction-limited observations provide higher spatial resolution 
than those coming from HST and the related astrometric measurements are thus potentially more precise.
This will be especially true in the future, with the advent of Extremely Large Telescopes (ELTs), which will
be larger by more than a factor of five than any future space telescope. In preparation for this future leap in telescope size available to
ground-based astrometry, it is important to explore the technical requirements that will lead to the full exploitation of MCAO
for PM measurements on an ELT.

Currently, the only operational MCAO facility
is the Gemini Multi-Conjugate Adaptive Optics System (GeMS, \citealt{rigaut14, neichel14a}) mounted at the Gemini-South telescope.
GeMS has been shown to be able to reach a good astrometric precision of $\sim0.2$ mas for bright
stars and exposure times exceeding one minute on single-epoch observations (\citealt{neichel14b}). 
However, GeMS performance on multi-epoch data, even if separated by only few hours, has turned out to be a more complicated matter. 
In general, several distortion effects contribute to move a star around on the detector of a MCAO-assisted camera (e.g. \citealt{trippe10}).
One of the biggest problems is the time variablity of the distortions which make it complex to calibrate or model them. 
This explains why only one previous PM study of a GC exists using MCAO data (\citealt{ortolani11}, 
using MAD observations of the cluster HP1). Moreover, in the particular case of GeMS, the distortions also have an extra component 
that varies quickly, possibly due to gravity flexure or to the movement of the AO-bench (\citealt{neichel14b, lu14, ammons14}).
This component makes PMs with GeMS even more complex to measure.

Here, we present the first multi-epoch astrometric study from our Gemini/GSAOI campaign targeting Galactic GCs using GeMS 
(\citealt{turri14, turri15, massari16}).
In this work, we present a method that is able to correct MCAO distortions by exploiting
previously measured distortion-free stellar positions and PMs. This allows us to test the GeMS 
astrometric performance for the first time on PM measurements obtained from multi-epoch data separated by few years. 
The initial test case that we present in this Letter is the GC NGC~6681, whose stars PMs have already been measured 
with HST by \cite{massari13} (hereafter Ma13).

The Letter is structured as follows. In Section \ref{method} we describe in detail the data analysis and the 
method we used to model GeMS distortions. In Section \ref{pm} we check our PMs for systematic 
errors and compare them to the previous HST measurements. Finally, we summarise our conclusions in Section \ref{concl}.

\section{Data analysis and camera distortions}\label{method}

When measuring stellar proper motions (PMs), it is fundamental to first disentangle the effect that distortions
have on the observed positions of the stars.

The GeMS facility is an extremely powerful 
instrument to investigate the kinematical properties of GCs via PMs, since it provides  
diffraction-limited observations across a large field of view (FoV) of $85$\arcsec$\times85$\arcsec with
a pixel scale of $0.02$ \arcsec pixel$^{-1}$.
However, GeMS suffers from distortions that vary significantly with time (\citealt{neichel14b}), due in part 
to variable seeing conditions and field-to-field differences in asterisms.
This makes it difficult to determine a stable, generic distortion model using previous GeMS observations
with already well established self-calibration techniques (e.g. \citealt{libralato14}). 
Moreover, it also means that each GeMS exposure requires its own distortion solution.
In the following, we provide a possible solution to this problem which relies on 
a-priori knowledge of distortion-free positions and accurate PMs for a sample of stars well distributed 
over the FoV. 

We selected GeMS images of the Galactic GC NGC~6681\footnote{Programme IDs: GS-2012B-SV-406, 
GS-2013A-Q-16, GS-2013B-Q-55, PI: McConnachie}, for which we have already measured accurate PMs in
the context of the HSTPROMO collaboration (Ma13). These frames are particularly interesting because 
they also include stars from the Sagittarius dwarf spheroidal galaxy (Sgr dSph). This will allow us to test 
the accuracy and precision on the relative proper motions between NGC~6681 and Sgr dSph.
The GeMS data consists of $8\times160$ s exposures in both the J and the Ks filters,
dithered by a few, non-integer pixel steps to cover the inter-chip gaps of the camera.

The pre-reduction of each raw image followed the procedure described in \cite{massari16}, where we
use both sky- and dome-flats for the flat-fielding. The photometry was then performed using DAOPHOT 
(\citealt{stetson87}). Each of the four chips of the camera
were treated separately. The PSF modelling involved fitting the light profile of a few hundred
bright, isolated and non-saturated stars with a Moffat function and allowing the residuals of such a fit 
to be described with a look-up table that varies cubically across the FoV.
The best-fit model has then been applied using ALLSTAR to all of the sources found $3\sigma$ above the
background, and each raw star position (x$_{i}^{r}$, y$_{i}^{r}$) was obtained as output. 

To obtain distortion-free positions (x$_{i}^{c}$, y$_{i}^{c}$), we considered that for evolved stars 
NGC~6681 has a central velocity dispersion of $\sigma_{0}=0.15$ \masyr that at the half-light
radius diminishes to $\sigma_{hr}=0.12$ \masyr (\citealt{watkins15}).
Stars belonging to the cluster Main-Sequence show even larger dispersion, up to $\sim0.2$ \masyr.
This means that stars with a PM error smaller than this value should be representative of the internal kinematics
of the cluster itself. To be conservative, we selected only
stars with errors smaller than $0.03$ \masyr that belong to the cluster according to their location
in the Vector Point Diagram (VPD) from the PM catalogue of Ma13. 
The positions of these $7770$ stars on the HST, distortion free reference 
frame adopted in Ma13 (already corrected for geometric distortions using the solution provided by
\citealt{anderson07} and aligned to Right Ascension and Declination by construction)
were then moved according to their PM to construct a distortion-free reference frame 
at the epoch of the GeMS dataset, $6.914$ years later. 
At this point, each chip of each GeMS exposure has been transformed onto this reference using a high-order polynomial 
(the best choice turned out to be a 5th order polynomial, see \citealt{massarispie} for details). 
Since the reference and GeMS exposures are at the same epoch, the stars should be aligned
and the polynomial only determines the distortion terms. The accuracy
achieved with this distortion correction is $\sim1$ mas across the majority of the FoV, degrading to $\sim2$ mas only in its corners
and near the edges (where less stars are available). This budget includes $\sim0.2$ mas due to the 
propagation of the PM uncertainty ($0.03$ \masyr) for the temporal baseline of $6.914$ yr. 

\begin{figure*}
    \includegraphics[width=\columnwidth]{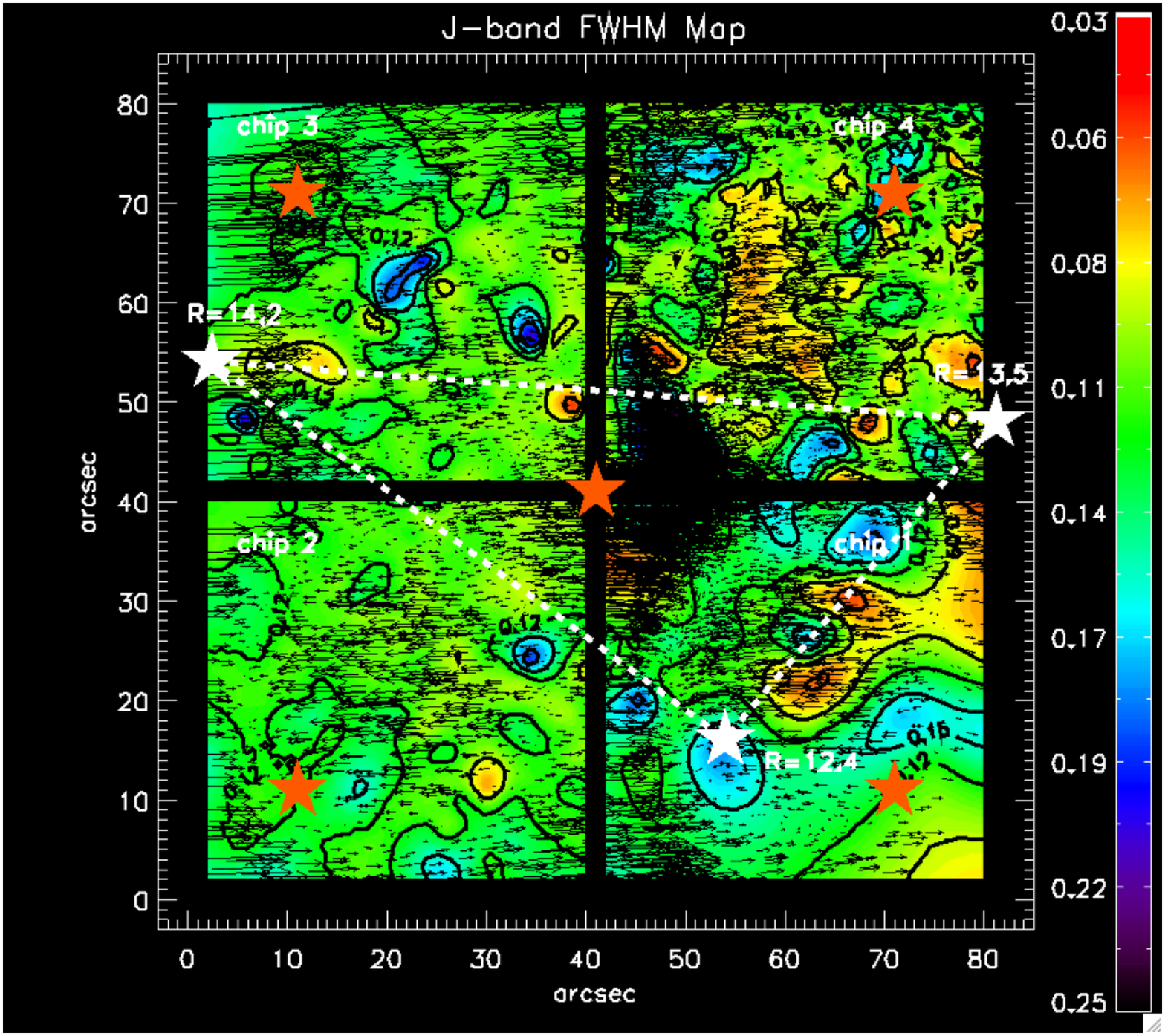}
    \includegraphics[width=\columnwidth]{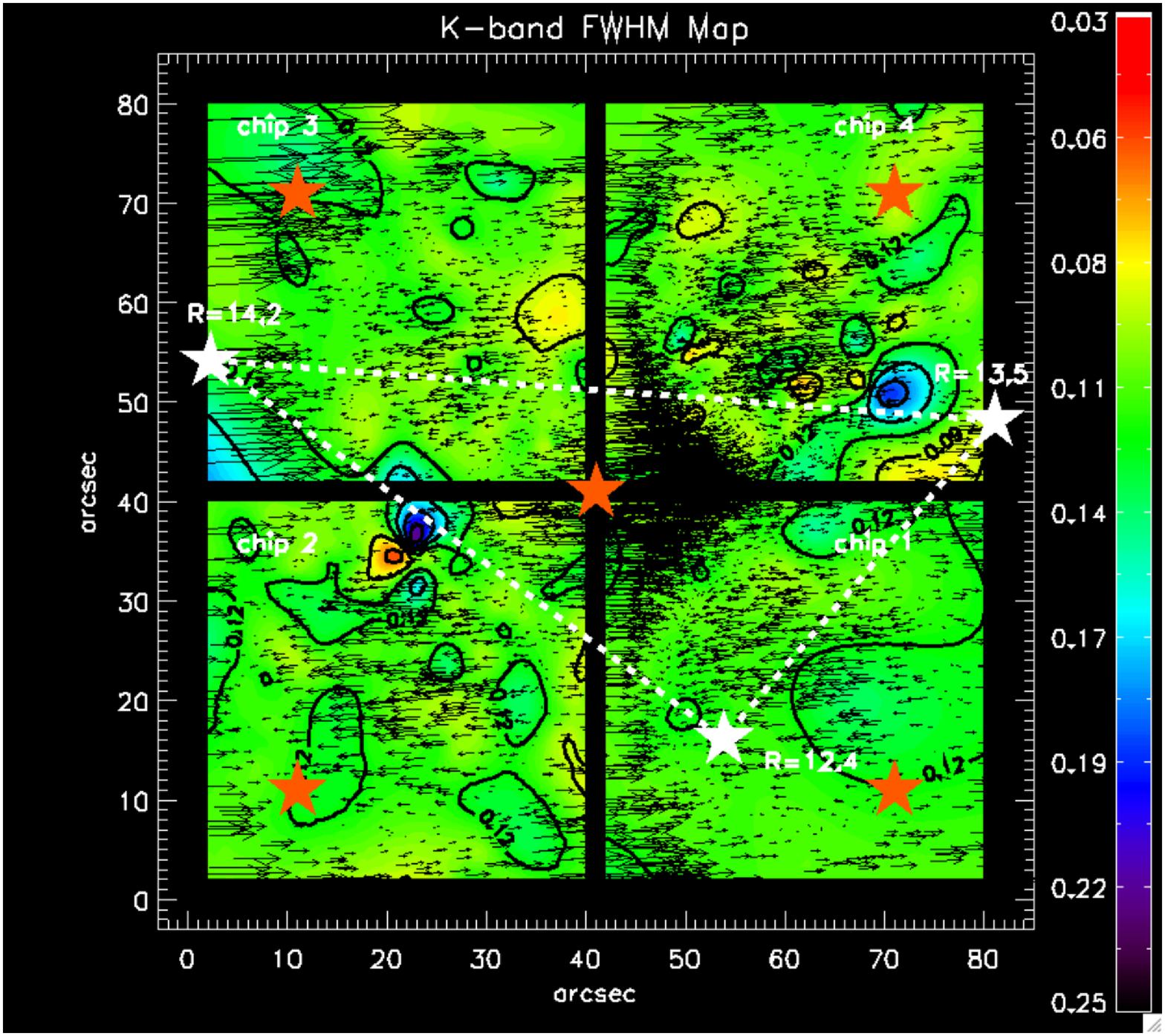}
        \caption{\small Distortion maps for the first exposure in the J (left panels) and Ks filter (right panels), overplotted
        to the corresponding FWHM maps. Vectors lengths are multiplied by a factor of 40 to highlight the distortion structures. 
        The adopted natural guide stars asterism is shown as a white triangle, while the LGS constellation is marked with orange 
        star-like symbols. The R magnitude of the natural guide stars is also labelled.}
        \label{dmaps}
\end{figure*}

To visualise how the GeMS distortions affect the position of stars across the FoV, in Fig.\ref{dmaps} we show an example of 
distortion map (computed in this case for the first of the available exposures) in the J and Ks filters, overplotted to the 
corresponding FWHM map. Each single exposure distortion map is the difference between the positions corrected with the 5-th order 
polynomial (x$_{i}^{c}$, y$_{i}^{c}$), and the positions corrected using only conformal linear transformations. 
We underline that in the upper-left corner of the chip 3, the polynomial solution is extrapolated since there are no stars in common 
with the HST FoV of Ma13, and might not be accurate. The distortions do not seem to follow the geometry described by the adopted asterism 
of the natural guide stars (white stars in Fig.\ref{dmaps}), nor that of the laser guide stars (orange stars in Fig.\ref{dmaps}). 
Also, no correlation is evident between the distortion pattern and that described by the FWHM (or Strehl ratio) maps, as the latter is well behaved 
and uniform across the entire FoV.

The following analysis refers to the mean distortion maps, computed by averaging the displacements vectors for all the stars common to 
all of the eight exposures per filter. We address the reader to \cite{massarispie} for their visualisation, but we underline that the 
general trend is very similar to that shown here in Fig.\ref{dmaps}.
Several structures in the distortions are seen in all of the chips and both filters.
The comparison between the coefficients of the 5th order polynomials used to model the J- and K-band distortions revealed that their 
mean difference is only $2\times10^{-7}$ (with a dispersion of $2\times10^{-6}$), thus demonstrating how strongly correlated the two solutions are.
The amplitude of the distortions in the X-component (corresponding to Right Ascension) is 
significantly larger than that in the Y-component (Declination). In fact the 
former spans an interval ranging from $-1.64$ pixels to $4.83$ pixels, while the latter varies only from $-0.49$ pixels
to $0.35$ pixels. In addition, a circular structure is seen roughly at the centre of each chip (which does not move from
exposure to exposure) where the X-component distortion is minimal and then changes its direction.
The same overall structure has been found using an independent method on GeMS observations of the GC NGC~1851 by S. Mark Ammons
(private communication, a detailed analysis of the distortions is beyond the aim of this paper and is provided in \citealt{massarispie}),
and consistent results were obtained in Dalessandro et al. submitted using GeMS observations of NGC~6624.
Finally, the average variation of the single-exposure maps with respect to the mean models shown in Fig.\ref{dmaps}
is $\sim0.05$ pixels, while the total variation ranges from $-1.1$ pixels to $1.3$ pixels. The exclusion of the corners
significantly limits the range between $-0.02$ to $0.03$ pixels.

Now, the distortion-corrected GeMS positions (x$_{i}^{c}$, y$_{i}^{c}$) can be used as second epoch and matched 
using linear transformations to the HST, first epoch positions, 
to measure the PMs for all of the sources that were not used to build the distortion-free reference frame.

\section{Proper Motions results}\label{pm}

Proper motions were measured following the procedure described in detail in Ma13.
Briefly, once all of the 8 first-epoch and 16 second-epoch exposures have been transformed onto the same reference 
frame as described in the previous section, we computed first- and second-epoch median positions adopting a 
3$\sigma$-rejection algorithm. Then, after discarding all of the sources with less than three single-epoch measurements, 
we computed the PM for each star as the displacement between its two median positions in $\Delta t=6.914$ yr.
We did not need to re-iterate the process of discarding non-member stars because the list of stars used to compute the transformations 
is made up of only NGC~6681 members by construction, since we used the information coming from the previous HST PM measurements. 
The PM errors were then computed as the sum in quadrature of each single epoch positional errors. In turn, these were defined as
the rms of the positional residuals about the median value, divided by the square root of the number of measurements.


\begin{figure}
    \includegraphics[width=\columnwidth]{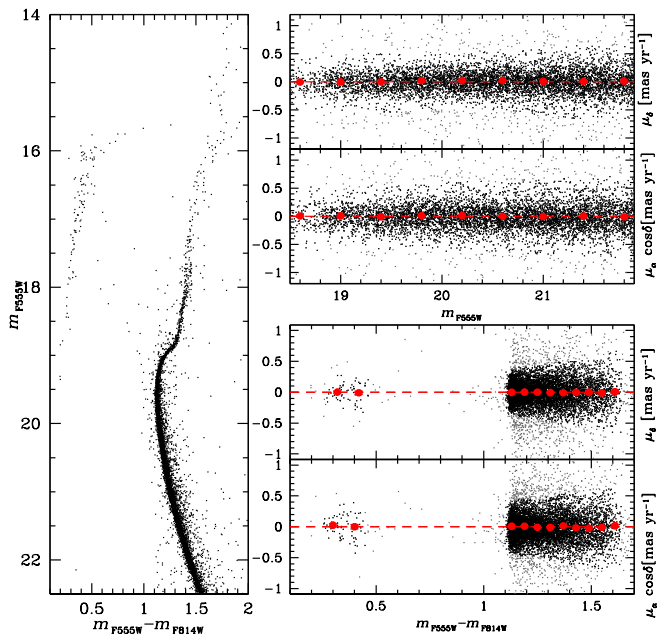}
        \caption{\small Systematic trends of the two components of PMs with {\it m}$_{{\rm F555W}}$-{\it m}$_{{\rm F814W}}$ colour 
        (bottom-right panels) and {\it m}$_{{\rm F555W}}$ magnitude (upper-right panels). PMs values rejected by the clipping algorithm are shown in grey. 
        In both cases, the results are consistent with no trend. The HST optical CMD from Ma13 is also shown in the left panel for the sake of comparison.}\label{system}
\end{figure}

Next, we performed several tests to check the consistency of our measurements. 
Following the analysis of systematic effects adopted in \cite{bellini14} and \cite{massari15}, we checked for
any possible trend in each PM component with star colours and magnitudes. The results of these tests are 
shown in the bottom- and upper- right panels of Fig.\ref{system}, respectively.
To check for any colour trends, we computed $3\sigma$-clipped mean PM values for bins of $0.06$ mag in ({\it m}$_{{\rm F555W}}$-{\it m}$_{{\rm F814W}}$) 
colour (taken from Ma13, see the corresponding CMD in the left panel of Fig.\ref{system}) 
and plotted them as red filled circles, with associated errorbars. 
All of the binned mean values turned out to be consistent with 0 within 1$\sigma$, 
thus excluding the presence of systematic errors with colour. We repeated the same procedure using {\it m}$_{{\rm F555W}}$ magnitude bins of 0.1 mag,
to find again that no trend exists between each PM component and stellar magnitude within $1\sigma$ uncertainty.

\begin{figure}
	\includegraphics[width=\columnwidth]{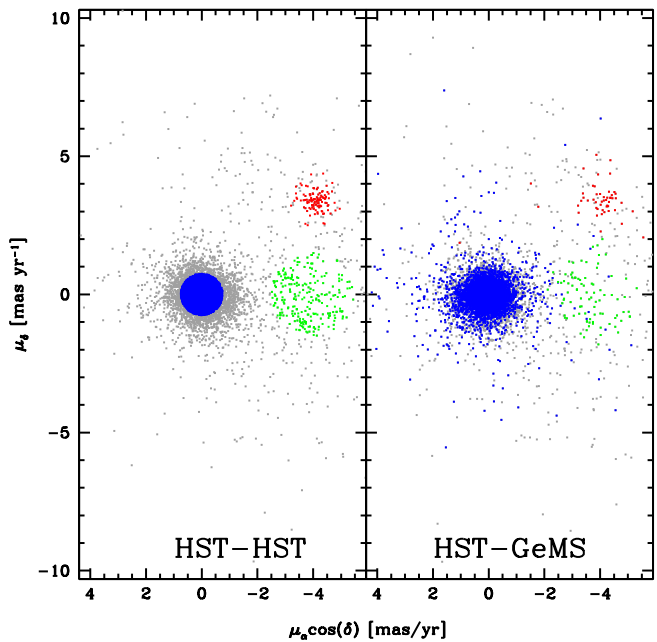}
        \caption{\small Comparison between the VPDs as obtained from Ma13 PMs and the PMs measured with the method proposed in this Letter.
        The kinematical distinction of the three populations (cluster stars in blue, Sgr dSph stars in red, field stars in green) is
        recovered on average. The few outliers are either very close to the image edges or are beyond the photometric regime of non-linearity.}\label{vpd}
\end{figure}

In Fig.\ref{vpd} we compared the VPDs as obtained from HST PMs (left panel, Ma13) and from this study (right panel).
Ma13 kinematically distinguished three different populations in the FoV of NGC~6681: the cluster (blue dots), the Sgr dSph (red dots) and the 
field, mostly composed of bulge stars (green dots). Plotting the PMs of the same stars in the two panels (keeping
the same colour code), we were able to retrieve the same separation with our new PMs. 
This is the first important finding, since only cluster stars have been used to build the distortion-free reference at the GeMS epoch, and no 
PM was given as input to the stars belonging either to the Sgr dSph or to the field. This means
that the method used is effective in correcting the distortions at least to the level of kinematically distinguishing the three populations.
By comparing the relative PM between NGC~6681 and the Sgr dSph, we can test for the first time
the accuracy of GeMS in providing a new epoch separated by several years for PMs measurement.

Ma13 found a PM of the cluster relative to the Sgr dSph of $(\mu_{\alpha}\cos\delta, \mu_{\delta}) = (4.12\pm0.03,-3.38\pm0.03)$ \masyr.
By using all of the Ma13 Sgr dSph stars that we also found in the GeMS exposures (red dots in Fig.\ref{vpd}), we found such a relative 
PM value to be of $(\mu_{\alpha}\cos\delta, \mu_{\delta}) = (4.09\pm0.06,-3.41\pm0.06)$ \masyr. This means that our newly found HST-GeMS PMs are as accurate
as those coming from HST observations within only 0.03 \masyr.

\begin{figure}
	\includegraphics[width=\columnwidth]{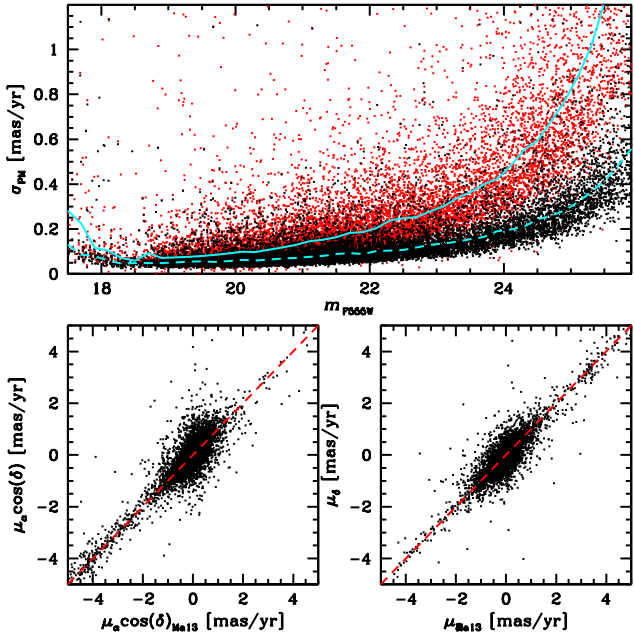}
        \caption{\small PM errors in the Right Ascension and Declination components as found combining two HST epochs (black dots) 
        or the corrected GeMS images with the first epoch of HST data (red dots). Cyan lines describe the 3$\sigma$-clipped mean
        trends in the two cases (continuous and dashed lines corresponding to the HST-HST and the HST-GeMS cases, respectively).
        From the comparison it is evident that the performance is similar until {\it m}$_{{\rm F555W}}\simeq21.5$ mag, where our new PM precision becomes
        about a factor of two larger than that coming from HST.}\label{err}
\end{figure}

The precision achieved is also comparable. The PM dispersions for Sgr dSph stars found with HST are 
$(\sigma_{PMX,Sgr}, \sigma_{PMY,Sgr}) = (0.23, 0.23)$ \masyr. This is given by the sum in quadrature between the PM errors
at the magnitude of Sgr dSph stars ($\sim0.16$ \masyr at {\it m}$_{{\rm F555W}}>22$ mag, see the upper panel of Fig.\ref{err}) and the 
Sgr dSph internal velocity dispersion ($\sim0.1$ \masyr, \citealt{frinchaboy12}). The dispersions found in this work are 
about a factor of two larger, being $(\sigma_{PMX,Sgr}, \sigma_{PMY,Sgr}) = (0.43, 0.43)$ \masyr. 
This is due to the larger PM errors ($\sim0.36$ \masyr at {\it m}$_{{\rm F555W}}>22$ mag, see the upper panel of Fig.\ref{err}) and to
the contribution coming from the error on the distortion correction, that as discussed in \cite{massarispie}, amounts to
$\sim0.14$ \masyr. The sum in quadrature of these two contributions and the Sgr dSph internal motions gives
the total dispersion. 
However, we underline that our PM precision becomes worse than that from HST only
for relatively faint stars ({\it m}$_{{\rm F555W}}>21.5$, i.e. the range where Sgr dSph stars fall), while the precisions are similar 
for brighter stars ($\sim0.05$ \masyr). This is clearly shown in the upper panel of Figure \ref{err}, where the cyan continuous and dashed lines 
describe the 3$\sigma$-clipped mean trends for the HST-HST and HST-GeMS PMs, respectively. 
Finally, the same dispersion quoted above is found also when computing the scatter around the 
1:1 relation that describes the comparison between the PMs of Ma13 and those measured in this work, shown in the bottom 
panels of Figure \ref{err}.

For sake of comparison, the only other previous PM measurements obtained with an MCAO
facility (\citealt{ortolani11}) reached a precision lower by a factor of $\sim2.5$, after combining their MAD observations
of the globular cluster HP1 (at a distance of $6.8$ kpc, that is $\sim3.2$ kpc closer than NGC~6681, \citealt{ferraro99}) with seeing-limited data obtained 
$14.25$ years before with the ESO New Technology Telescope.
In that case the authors did not correct the MAD data for the camera distortions, but estimated a contribution $\sim 0.2$ \masyr.
This value is significantly smaller than the error term that would come from GeMS distortions without their correction ($\sim4$ \masyr,
see \citealt{massarispie}). One difference between GeMS and MAD is that the former is Cassegrain mounted, while the latter is mounted at Nasmyth. 
Therefore it is plausible that MAD dataset suffered from smaller distortions, but since a detailed treatment
of MAD distortions is not available in that paper, we cannot exclude the possibility that they have been underestimated. This statement
is further supported by the findings of \cite{meyer11}, where the authors estimated the distortions affecting MAD observations
of the GC NGC6388 to be in the range from $\sim3$ to $\sim40$ mas. A further investigation on MAD distortions found corner to corner values
of $\sim100$ mas (A. Bellini, unpublished work).

\section{Conclusions}\label{concl}

In this work we present the first PM measurements obtained exploiting the potential of the MCAO GeMS/GSAOI camera. 
We used the a-priori knowledge of independently measured PMs for stars in the FoV of interest
to model the distortions affecting the camera, finding a different distortion solution for each of the GeMS exposures.
After combining the corrected GeMS images with observations coming from HST, providing a temporal baseline of $\Delta t=6.914$ yr,
we were able to determine the relative PM between the globular cluster NGC~6681 and the Sgr dSph galaxy. The value we found of
$(\mu_{\alpha}\cos\delta, \mu_{\delta}) = (4.09,-3.41)$ \masyr matches the previous HST measurement with an 
accuracy of $0.03$ \masyr. Also the achieved precision turns out to be comparable to that of HST, 
and become worse by a factor of two at the faint magnitudes of Sgr dSph stars.

Our findings demonstrate that GeMS is potentially useful to obtain high-quality PMs from the ground,
when combined with space-based observations.
We are waiting for two GeMS epochs sufficiently distant in time to test the astrometric performances achievable using only MCAO datasets.
This work sets the stage for future PM measurements with this complex AO instrumentation, providing strong support for the need of a 
careful treatment of time-varying distortions.

Following the effective investigation discussed in this Letter, our group is now testing a new observational strategy to use GeMS for PM
measurements without the support of previous HST data.
This involves the combined use of the FLAMINGOS-2 (\citealt{eiken04}) and GeMS/GSAOI images. According to our strategy,  
distortion-corrected FLAMINGOS-2 seeing-limited observations will precede any GeMS dataset to provide an independent reference 
frame to be used as a corrector for the distortions of each GeMS exposure. 
In order to make the above procedure possible, we have obtained dedicated time to model the FLAMINGOS-2 camera 
distortions (PI A. McConnachie). 
The proposed strategy will pave the road for future astrometric MCAO observations, including those with ELTs.

\begin{acknowledgements}

We thank the anonymous referee for her/his comments and suggestions which helped us to improve the presentation of our results.
Based on observations obtained at the Gemini Observatory. Acquired through the Gemini Science Archive and processed using the Gemini IRAF package.
DM and GF have been supported by the FIRB 2013 (MIUR grant RBFR13J716).

\end{acknowledgements}

%
\bibliographystyle{aa} 
\bibliography{ms.bib}




\end{document}